\documentclass[aps,pre,reprint, amsmath, amssymb, superscriptaddress]{revtex4-2}

\usepackage{bm}
\usepackage[retainorgcmds]{IEEEtrantools}
\usepackage{graphicx}
\usepackage{framed}
\usepackage{placeins}
\usepackage{mathrsfs}
\usepackage{amsmath}
\usepackage{amssymb}
\usepackage{color}
\usepackage{amsfonts}
\usepackage{amsthm}
\usepackage{times,txfonts}
\usepackage{enumerate}
\usepackage{dsfont}
\usepackage{nicefrac}
\newlength{\eqboxstorage}

\begin{document}

\title{Emergence of opinion splits in the Sznajd model with latency}
\date{\today}
\author{Ryan W. Salatti}
\affiliation{Universidade Federal do ABC,  09210-580 Santo Andr\'e, Brazil}
\author{Andr\'e M. Timpanaro}
\email{a.timpanaro@ufabc.edu.br}
\affiliation{Universidade Federal do ABC,  09210-580 Santo Andr\'e, Brazil}
\begin{abstract}
In the modelling of social systems, opinion latency is the idea that once an agent changes its opinion, there will be a period of time where it is immune to other changes. When added to the voter model this leads to a situation where no matter how low the latency is or how many opinions are considered, all opinions end up in a coexistence where they are equally represented. In this work, we examine what happens when latency is added to the Sznajd model. What we find is that for low latency, the model behaves roughly like it does in the absence of latency, where one opinion will always eventually dominate. For high latency, the possibility for a symmetric coexistence of 2 opinions arises, but contrary to the voter model, a coexistence of more than 2 opinions is never stable. We provide evidence of this phenomenon with computer simulations of the model in Barabási-Albert networks, together with a mean field treatment that is able to capture the observed behavior. We argue that this could hint at an explanation for the prevalence of two opinion splits in the real world.
\end{abstract}
\maketitle{}

\section{Introduction}

Inspired by spin models, binary opinion models are a popular tool in statistical physics for the modeling of the evolution of opinions in a society \cite{voter-review-2019, social-models-review-2015, analise-eleicoes-aguiar, analise-eleicoes-lituania}. In these models, society is represented by a network of agents that can have either opinion $A$ or opinion $B$, with the time evolution of these opinions being a consequence of interactions between the agents.

In the long time limit, these models can either converge to a state where all agents share the same opinion (which we will call a consensus state) or to a state where agents retain different opinions (a polarized state). In their simplest form, these models typically tend to consensus \cite{voter-def, sznajd-def, guf-def, qvot-def}, but since their inception, there have been important modifications that allow for the coexistence of different points of view, such as the addition of random opinion changes (noise) \cite{noisy-voter, Timpanaro-Sznajd-2011}, the addition of agents that don't change their opinion \cite{zealots}, generalizations to opinions lying in a continuum between 2 extremes with interaction requiring close enough opinions \cite{deffuant-def, HK-def}, aging effects \cite{aging-eguiluz, aging-peralta, aging-toral}, and the addition of latency periods, where an agent can't change its opinion too frequently \cite{lambiotte-latency, cimini-latency}.

The importance of opinion models where a polarized state can exist is due to how frequent this polarization is in the real world, with political polarization being the most obvious example. Also, in many real world examples of polarization, there exists a spectrum between two extremes. On one hand, binary opinion models are a good choice for modeling this type of polarization, since it includes the existence of two extremes since the beginning. But on the other hand, they don't provide any clarification on why these types of situations are so common.

In this work we explore the possibility of using models with more than two opinions as a way to explain why two opinion splits are so prevalent. Our focus will be in the Sznajd model \cite{sznajd-def} with the addition of latency, but we will also make some comparisons with the voter model with latency \cite{voter-def, lambiotte-latency}.

The addition of latency can be thought of as splitting each opinion state $\sigma$ into two: $L_{\sigma}$ (latent with opinion $\sigma$) and $A_{\sigma}$ (active with opinion $\sigma$). Once an agent changes opinion to $\sigma$ it goes into the $L_{\sigma}$ state. While it remains in this state, any attempts to further change its opinion fail (or have a reduced probability of success in some variants \cite{sci-rep-latency, prl-latency}). Depending on the variant of latency used, an agent in state $L_{\sigma}$ can switch to state $A_{\sigma}$ with either some constant rate \cite{lambiotte-latency} or after some fixed amount of time \cite{cimini-latency}. Once the agent is in the state $A_{\sigma}$ it can freely change its opinion again (but once it does change it becomes latent and the cycle repeats).

This type of modification aims to model the idea that once we make a decision that changes our position, this comes with some kind of cost that makes it less likely that we switch our opinion too quickly. It has been explored mostly in the context of the voter model and always in the case of two opinions, where it has been successfuly used to describe polarization \cite{lambiotte-latency} and opinion oscilations \cite{cimini-latency}.

In this work we propose a way this modification can be introduced in the Sznajd model and investigate the behaviour of both the Sznajd model and the voter model with more than two opinions and with the addition of latency (more precisely, the variant where the agents become active again with a constant rate). Also, in order to capture some properties of real world social networks, we will use a scale-free (Barabási-Albert) network as the model for the social network of the agents.

Our results are as follows:
\begin{itemize}
\item In the Sznajd model with latency, there exists a possibility of stable coexistence between 2 opinions, however, coexistences with more than 2 opinions are not stable. Moreover, there is a transition back to the usual consensus behaviour of the model when the latency becomes too low.
\item In the voter model, no matter how low the latency is, a symmetric coexistence of all opinions is the only stable situation.
\item This behaviour for both the Sznajd and voter models can be explained qualitatively by a mean field theory (which is equivalent to simulating the models in a complete network in the limit of the network becoming infinite).
\end{itemize}
It is worth noting that the behaviour of the Sznajd model with latency is in sharp contrast with what happens if you add noise to the Sznajd model with $M>2$ opinions, in which case we either have one majority opinion or a coexistence of all $M$ opinions having the same proportion of agents \cite{Maycon}.

The paper is organized as follows: In section \ref{sec:models} we present the precise description of the models employed. In section \ref{sec:sims} we present our simulation results for the Sznajd and voter models using Barabási-Albert networks. In section \ref{sec:MF} we present the mean field treatment we developed and summarize the results obtained, with the detailed calculations relegated to the appendices. Finally, we summarize our conclusions and present some discussions in section \ref{sec:conc}.

\section{Model Descriptions}
\label{sec:models}

As it is usual in opinion propagation models, the society will be represented by an undirected graph where each of the vertices represents an agent that has an opinion (or a state more generally) and an edge between two vertices represents a social connection between the corresponding agents. In these model descriptions, we will denote the neighbourhood of a vertex $i$ by $\Gamma_i$.

\subsection{The voter model with latency}
In \cite{lambiotte-latency} the voter model with latency is defined as

\vspace{0.2cm}
%\begin{framed}
\noindent{\bf Voter model with latency:}
\begin{itemize}
    \item At each timestep choose an agent $i$ drawn at random.
    \item If $i$ is latent, then with probability $p$ it becomes active. We move on to the next timestep.
    \item Otherwise if $i$ is active, we choose a neighbouring agent $j$ at random from $\Gamma_i$. If $i$ and $j$ have different opinions, then $i$ copies $j$'s opinion and becomes latent.
\end{itemize}
%\end{framed}
Note that even though this model has been mostly considered only in the case of two opinions, nothing really changes in the description if we have more than 2 possible opinions. An important point of this type of latency is that the agent being copied/convincing ($j$) doesn't need to be active to have its opinion copied. In other words, activity/latency is a completely separate property than the agent opinion.

\subsection{Our extension of the Sznajd model to include latency}

In order to extend the idea of latency to the Sznajd model we recall that the main difference between the voter and Sznajd models is that in the voter model only one agent is enough to change an opinion, while in the Sznajd model we need a pair of agents that agree in order for an opinion change to be possible. The idea behind this is that a group of people has more convincing power than isolated individuals and can also be thought of as a way of introducing \emph{peer pressure} into opinion dynamics. Two possibilities for this extension are

\vspace{0.2cm}
%\begin{framed}
\noindent{\bf Inflow Sznajd model with latency:}
\begin{itemize}
    \item At each timestep choose an agent $i$ drawn at random.
    \item If $i$ is latent, then with probability $p$ it becomes active. We move on to the next timestep.
    \item Otherwise if $i$ is active, we choose neighbouring agents $j$, drawn at random from $\Gamma_i$ and $k$ drawn at random from $\Gamma_j \setminus\{i\}$.
    \item If $j$ and $k$ have the same opinion while $i$ has an opinion that is different from the opinion of $j$ and $k$, then $i$ copies their opinion and becomes latent.
\end{itemize}
%\end{framed}
Like in the case of the voter model, $j$ and $k$ don't need to be active to convince $i$. In fact, they don't even need to be in the same state, as long as they have the same opinion.

This version is a fairly clear generalization of the rules for the voter model, but they are a bit different from the usual Sznajd model, as here there is an information inflow (the agent we choose at the start, copies the opinion of someone else) instead of an outflow (the agent chosen at the start convinces someone else) like in the usual Sznajd model. A possible way of extending the Sznajd model to have latency and still have an information outflow is as follows

\vspace{0.2cm}
%\begin{framed}
\noindent{\bf Outflow Sznajd model with latency:}
\begin{itemize}
    \item At each timestep choose an agent $i$ drawn at random.
    \item With probability $p$ we set the agent $i$ as active (if $i$ is already active, nothing happens). We move on to the next timestep.
    \item Otherwise (with probability $1-p$), we choose neighbouring agents $j$, drawn at random from $\Gamma_i$ and $k$ drawn at random from $\Gamma_j \setminus\{i\}$.
    \item If $i$ and $j$ have the same opinion while $k$ is active and has an opinion that is different from the opinion of $i$ and $j$, then $k$ copies their opinion and becomes latent.
\end{itemize}
%\end{framed}
As will be shown with the simulation results there are only minor differences between the behaviours of these two versions when a Barabási-Albert network is used. For the mean field treatment, the only difference is the timescale of the evolution which is slower for the outflow version.

\section{Simulation Results}
\label{sec:sims}

For our simulations, we used Barabási-Albert networks with $10^4$ agents and minimum coordination 3 (different networks were used, but always with these same parameters). In all simulations, time is measured in Monte Carlo Timesteps (MCTs) which is a number of timesteps equal to the number of agents (so in our case 1 MCT = $10^4$ timesteps). We also always used for the initial conditions the hypothesis that all agents start out as active. The simulations are aimed at understanding in which conditions coexistence of opinions is possible in the Sznajd model with latency and to a lesser extent when that is possible in the voter model with latency. Firstly, we show that situations with more than 3 opinions are unstable in the Sznajd model, showing that symmetrical coexistences with $M>2$ opinions ``decay'' into either coexistences of only two opinions or into consensus states. After that, we examine in more detail the case with only two opinions, checking which initial conditions lead to consensus and which lead to coexistence as the parameter $p$ is changed. Finally, we show that the voter model always ends up in a symmetric coexistence of all $M$ opinions, even if we start with heavily asymmetric initial conditions.

\subsection{Sznajd model with $M>2$}
\label{ssec:instability}

\begin{widetext}

\begin{figure}[ht!]
    \centering
    \includegraphics[width=0.29\textwidth]{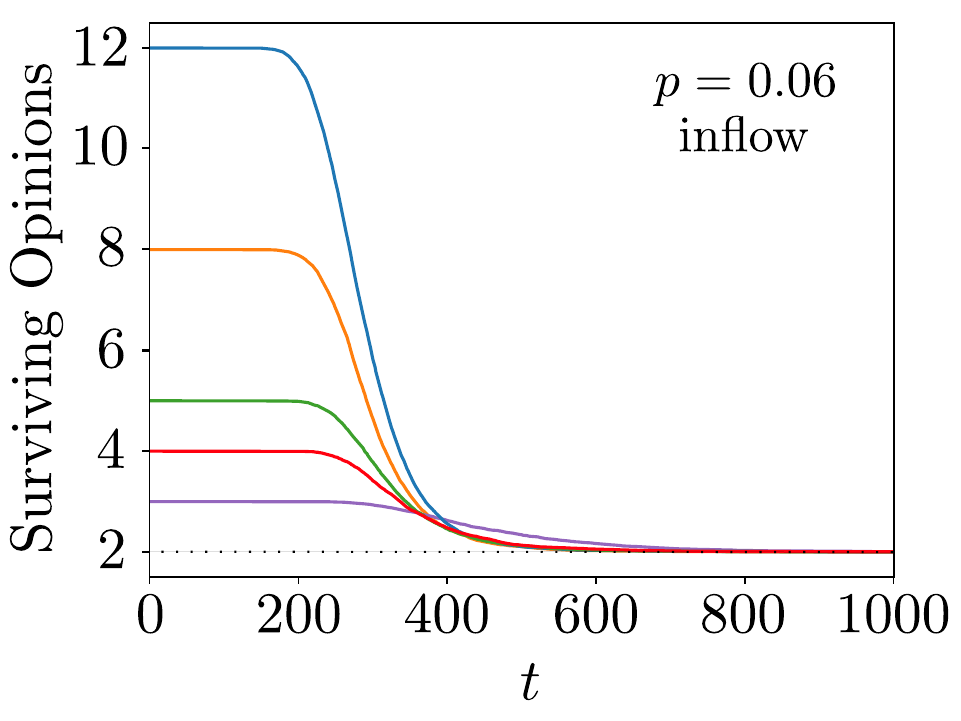} %0.06
    \quad
    \includegraphics[width=0.29\textwidth]{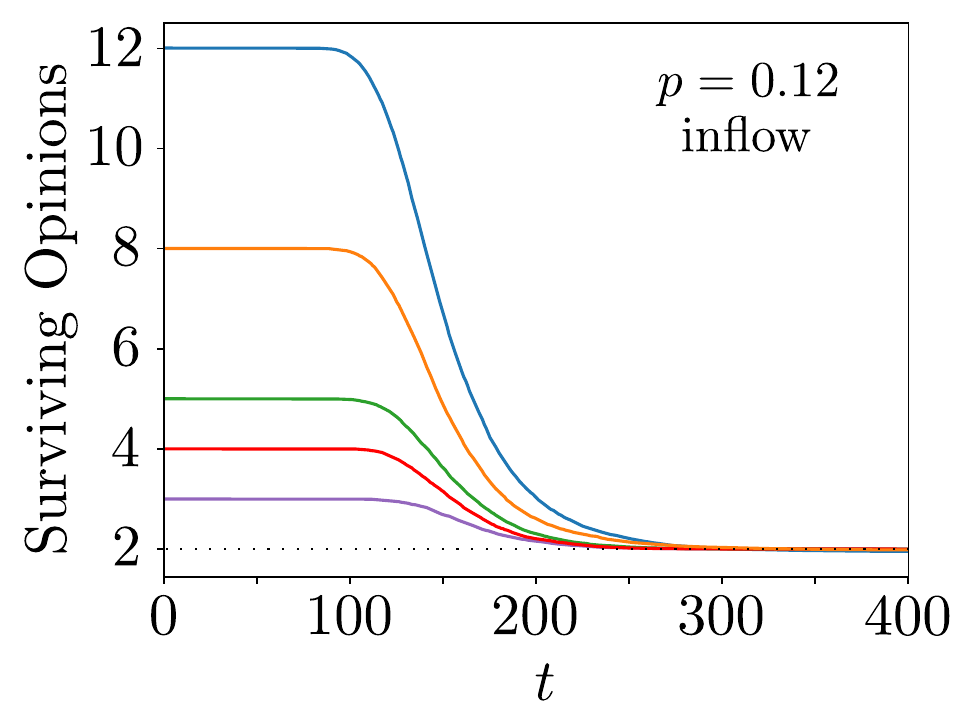} %0.12
    \quad
    \includegraphics[width=0.29\textwidth]{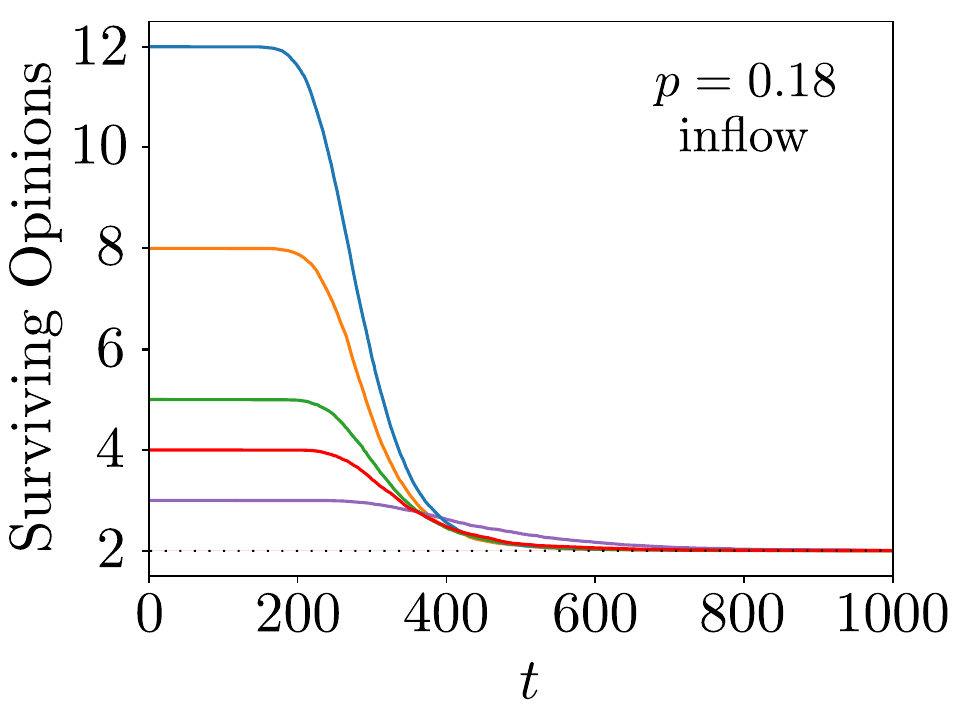} %0.18
%    \caption{Opinion Survival - inflow}
%    \label{fig:inflow-surv}
%\end{figure}
%
%\begin{figure}[ht!]
%    \centering
    \includegraphics[width=0.29\textwidth]{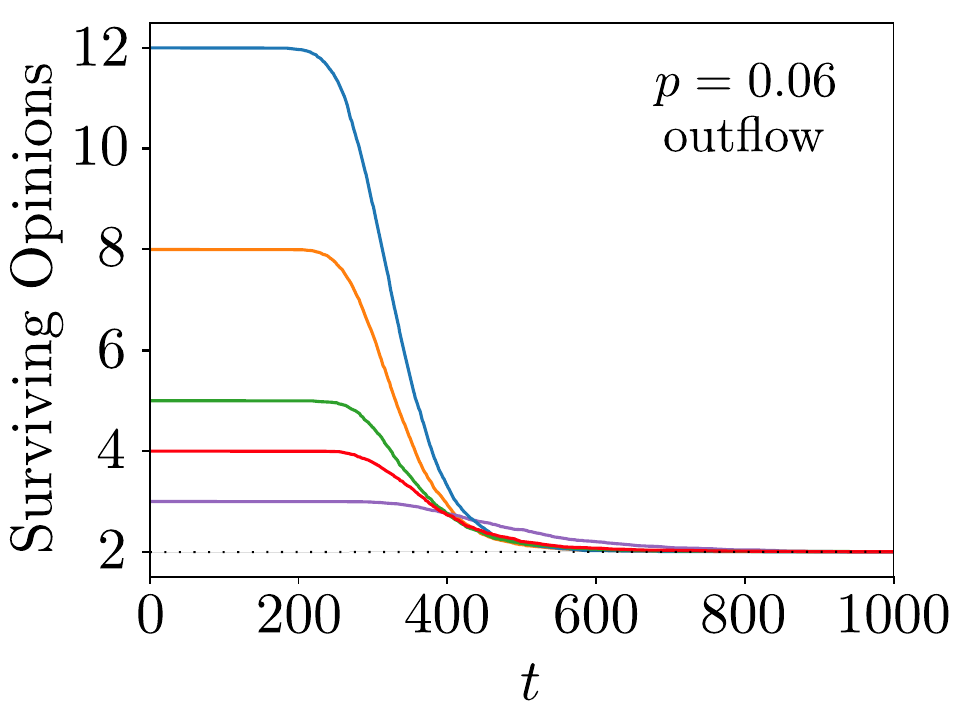} %0.06
    \quad
    \includegraphics[width=0.29\textwidth]{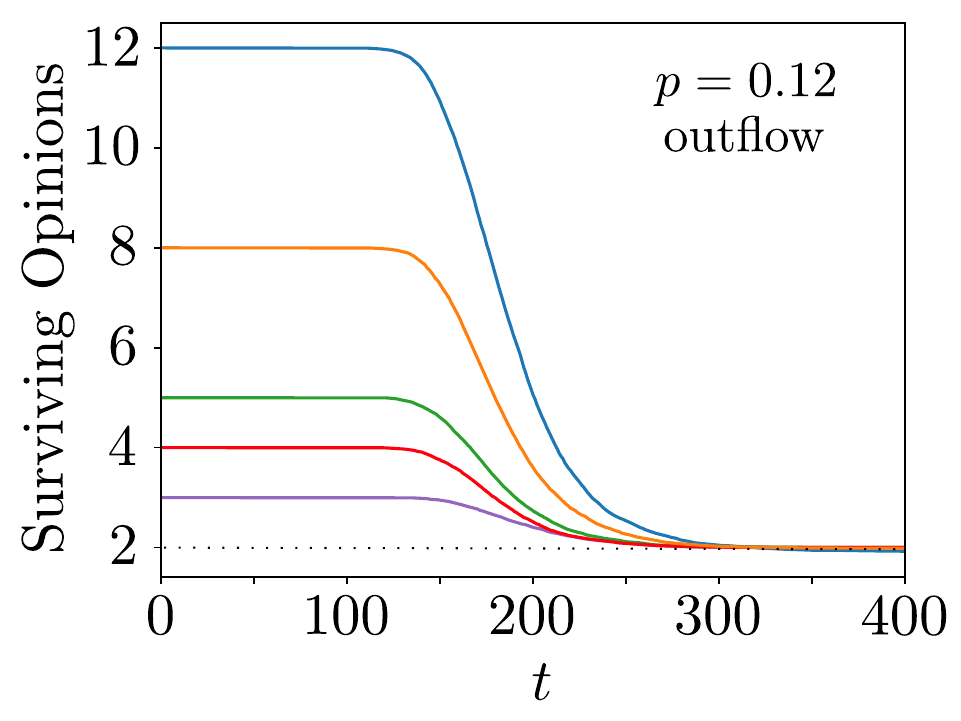} %0.12
    \quad
    \includegraphics[width=0.29\textwidth]{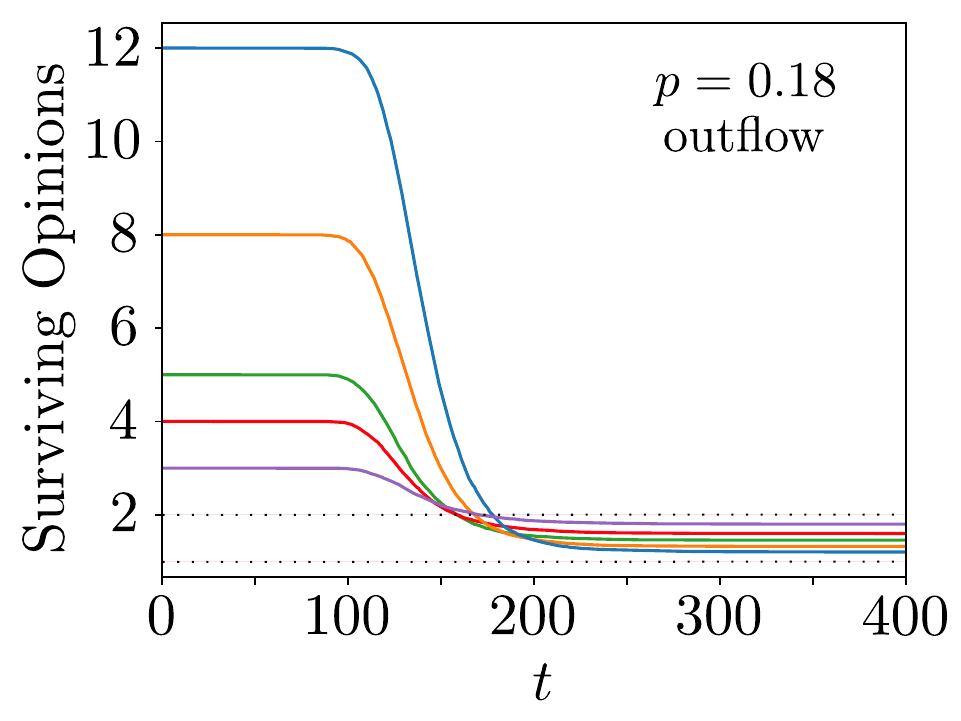} %0.18
    \caption{Average over $10^3$ simulations of the number of opinions present in the network as a function of time (measured in MCTs). The simulations were done in both versions of the Sznajd model with latency (inflow and outflow), for starting amounts of opinion $M_0 = 3,4,5,8, 12$ and $p$ ranging from $0.03$ to $0.3$ ($0.06, 0.12$ and $0.18$ shown in the figure).}
    \label{fig:survival}
\end{figure}

\end{widetext}

In figure \ref{fig:survival} we can see the average over $10^3$ simulations of the number of opinions present in the network as time goes by. These simulations were made for initial amounts of opinion $M_0 = 3, 4, 5, 8, 12$. The initial conditions are such that each agent gets assigned one of the $M$ opinions uniformly. In the simulations we used $p$ ranging from $0.03$ to $0.3$. The graphs in figure \ref{fig:survival} show the results for 0.06, 0.12 and 0.18. We see that the average tends to a value close to 2, meaning most of the simulations are ending up as a coexistence of two opinions, with the exception for the case $p = 0.18$ for the outflow model, where the averages end up between 1 and 2, so part of the simulations also end up in a consensus state.

\begin{figure}[htbp!]
    \centering
    \includegraphics[width=0.29\textwidth]{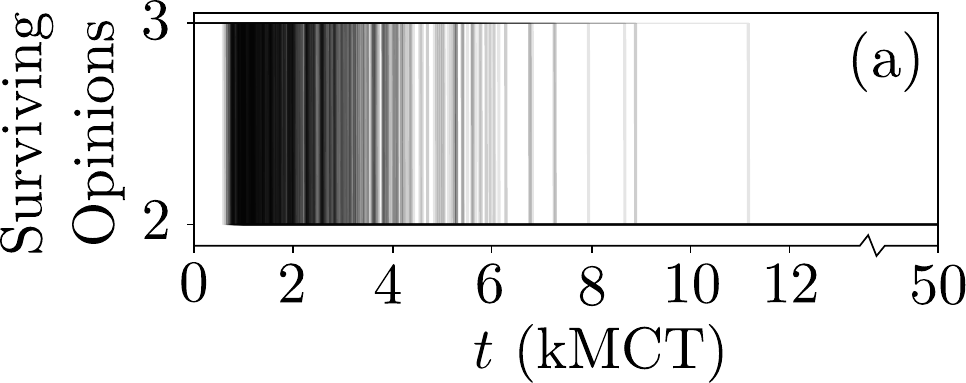} \includegraphics[width=0.29\textwidth]{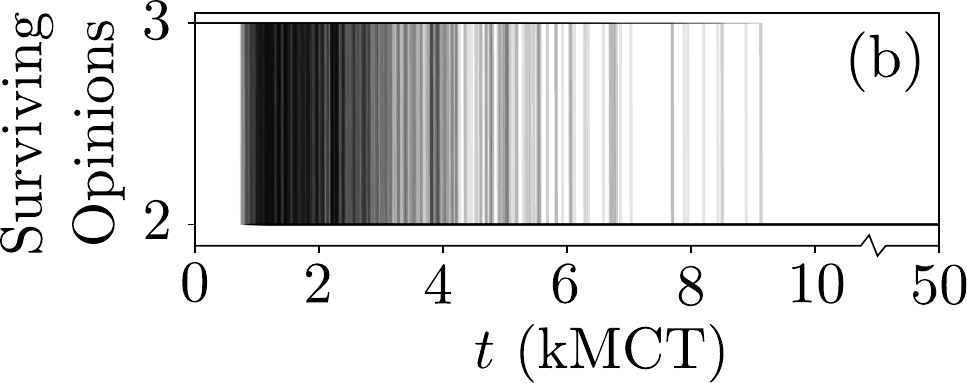}
    \caption{Timeseries of the amount of opinions present in the network for $10^3$ different simulations, starting with 3 different opinions and using $p=0.03$. For the time unit, 1 kMCT = $10^3$ MCTs. (a) Inflow Sznajd model. (b) Outflow Sznajd model.}
    \label{fig:M-3-surv}
\end{figure}
In order to exclude the possibility that some simulations are still finishing with more than 2 opinions, figure \ref{fig:M-3-surv} shows the timeseries of the opinion count for $10^3$ simulations, for the highest value of latency we examined ($p = 0.03$), which is where the longest lived states with 3 coexisting opinions were found. We can see that after around $10^4$ MCTs, all simulations decayed to having two opinions (while none ended up in a consensus state).

\subsection{Sznajd model with $M=2$}
The simulations so far establish that stable coexistences with 2 opinions are possible, but also that sometimes these simulations still evolve towards consensus, especially for larger values of the parameter $p$ (corresponding to lower latency, since latent agents turn active faster).

To understand this better, we turn our attention to how different initial conditions evolve in time for different values of the parameter $p$. For these simulations, we draw the initial condition such that each agent will start with opinion $\sigma$ with a probability $\theta_{\sigma\circ}$ and all initial opinions are drawn independently.

\begin{figure}
    \centering
    \includegraphics[width=0.3\textwidth]{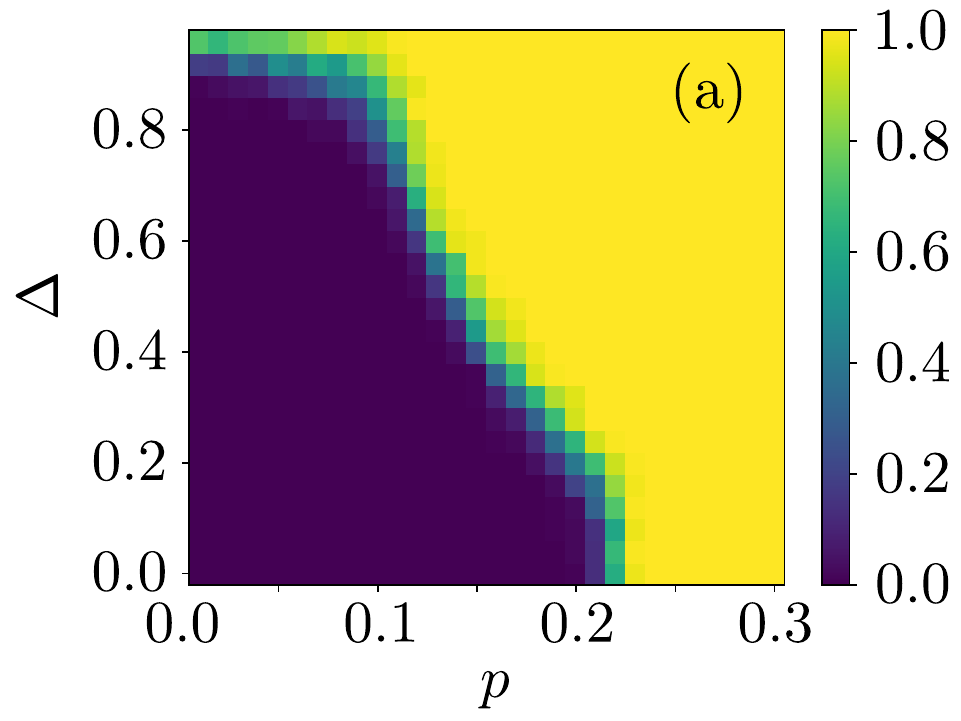} \includegraphics[width=0.3\textwidth]{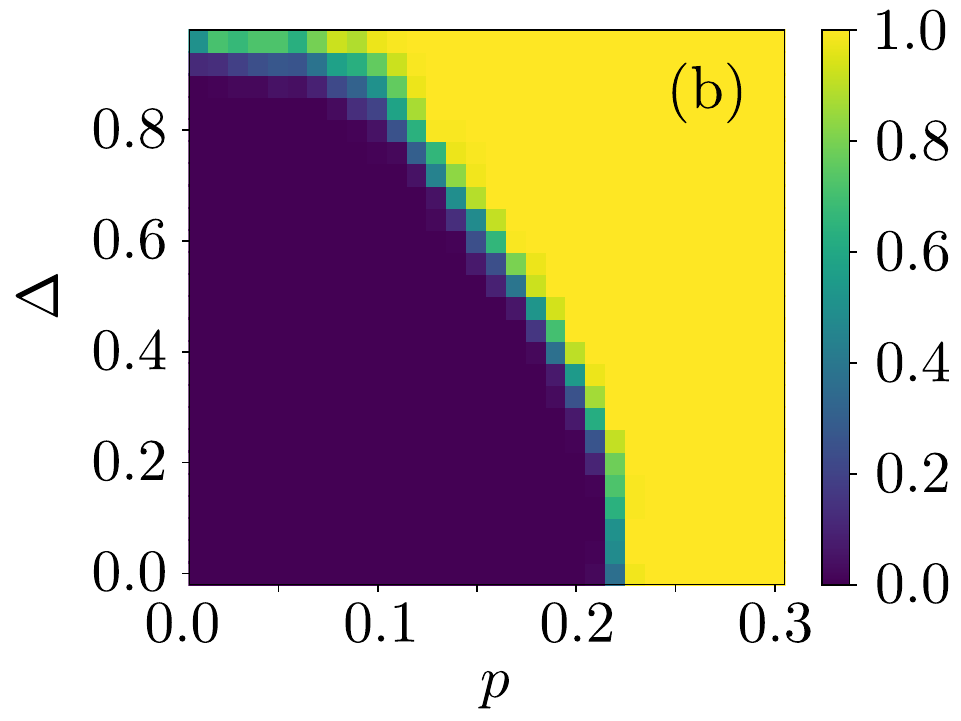}
    \caption{Proportion of simulations that ended up in a consensus state, starting with 2 different opinions, as a function of the parameter $p$ and of the asymmetry $\Delta = |\theta_{1\circ} - \theta_{2\circ}|$, where $\theta_{1(2)\circ}$ is the starting proportion of agents with opinion 1 (2). For each point in the graph, 100 simulations were made with a duration of $10^3$ MCTs. (a) Inflow Sznajd model. (b) Outflow Sznajd model.}
    \label{fig:sims-bacia}
\end{figure}

A graph summing up these results is found in figure \ref{fig:sims-bacia}, where we can see that both increasing $p$ or increasing the asymmetry $\Delta = |\theta_{1\circ} - \theta_{2\circ}|$ in the initial condition contribute to the model ending up in a consensus state. We can see that the outflow model tends to hold coexistences a little more, but the results are still very similar.

\subsection{Comparisons with the voter model}

\begin{figure}[htbp]
    \centering
    \includegraphics[width=0.4\textwidth]{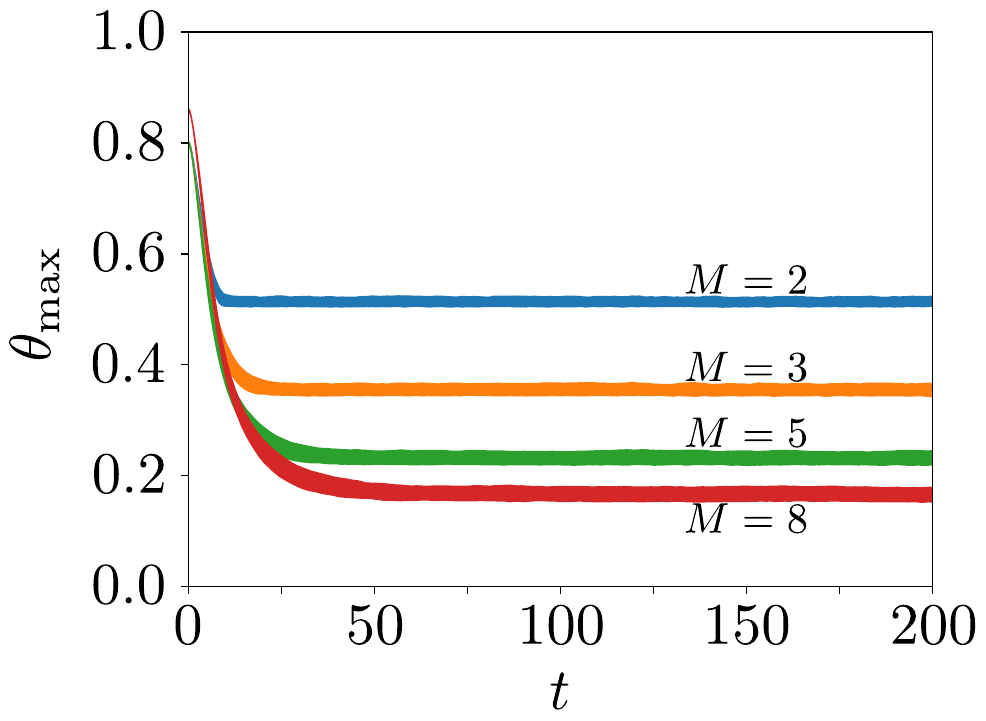} 
    \caption{Evolution of $\theta_{\mathrm{max}}$, the proportion of agents holding the most common opinion as a function of time. Simulations were made for the voter model with latency using $p=1$ and different initial amounts of opinions $M=2,3,5,8$. Denoting the starting proportion of agents holding opinion $\sigma$ by $\theta_{\sigma}$, the exact initial conditions were: $\theta_1 = 0.8$, $\theta_2 = 0.2$ for $M=2$; $\theta_1 = 0.8$, $\theta_2 = \theta_3 = 0.1$ for $M=3$; $\theta_1 = 0.8$, $\theta_2 = \ldots = \theta_5 = 0.05$ for $M=5$ and $\theta_1 = 0.86$, $\theta_2 = \ldots = \theta_8 = 0.02$ for $M=8$. Each band covers from one standard deviation below the average $\theta_{\max}$ to one standard deviation above the average. For each value of $M$, $10^3$ simulations were made. After 500 MCTs, all of the simulations still had all $M$ opinions present.}
    \label{fig:voter-surv}
\end{figure}

Finally, we turn our attention to the voter model. We want to show that contrary to the Sznajd model, a coexistence of all opinions is the only stable solution. In order to see this, we made simulations starting from very asymmetric conditions and using low latency ($p=1$), showing they converge to a symmetric coexistence. In order to do that, we look at the proportion of agents holding the most common opinion (we will denote it $\theta_{\mathrm{max}}$). We looked at $10^3$ simulations and the results can be found in figure \ref{fig:voter-surv}, where we see that starting with large $\theta_{\mathrm{max}}$ (and hence a high asymmetry in the initial condition), $\theta_{\mathrm{max}}$ drops to approximately $\nicefrac{1}{M}$ (indicating a coexistence). Also, after 500 MCTs none of the simulations done had lost any of their opinions.

%\FloatBarrier
\section{Mean Field Treatment}
\label{sec:MF}

We can develop a mean field treatment for these models if we consider them in a complete network with $N\gg 1$ agents. When we have a complete network, the models we want to describe become Markov chains with a state given by the amount of latent and active agents for each of the opinions. Furthermore, the dynamics will be such that we can easily write equations for the time evolution of the probabilities of a random agent having opinion $\sigma$ and being latent or active.

We will denote the probability that a random agent is latent with opinion $\sigma$ by $\eta_{\sigma}$ and the probability that it is active with opinion $\sigma$ by $\nu_{\sigma}$. Since the network is complete and $N \gg 1$, we can also assume that during the dynamics, all sites are randomly drawn from the whole network and neglect the possibility of any repetitions occurring.

Throughout our analysis, we will assume $p\neq 0$ because otherwise all the agents quickly become latent and the dynamics freeze completely.

\subsection{Sznajd Model}

Adding up all possibilities, one time step of the inflow Sznajd model would lead to

\begin{equation}
\left\{
\begin{aligned}
& \Delta \eta_{\sigma} = \frac{1}{N}\left(-p \eta_{\sigma} + (\eta_{\sigma} + \nu_{\sigma})^2 \sum_{\sigma'\neq \sigma} \nu_{\sigma'}\right) \\
& \Delta \nu_{\sigma} = \frac{1}{N}\left(p\eta_{\sigma} - \nu_{\sigma}\sum_{\sigma'\neq\sigma} (\eta_{\sigma'} + \nu_{\sigma'})^2\right)
\end{aligned}
\right.
\label{eq:sznajdin-MF-bigraw}
\end{equation}
where the first terms in both equations come from latent agents becoming active and the remaining terms come from an active agent copying a new opinion from an agreeing pair and becoming latent. Since $N \gg 1$, if we take 1 MCT as the time unit, then the system (\ref{eq:sznajdin-MF-bigraw}) reduces to the system of differential equations

\begin{equation}
\left\{
\begin{aligned}
& \dot{\eta}_{\sigma} = -p \eta_{\sigma} + (\eta_{\sigma} + \nu_{\sigma})^2 \sum_{\sigma'\neq \sigma} \nu_{\sigma'} \\
& \dot{\nu}_{\sigma} = p\eta_{\sigma} - \nu_{\sigma}\sum_{\sigma'\neq\sigma} (\eta_{\sigma'} + \nu_{\sigma'})^2
\end{aligned}
\right.
\label{eq:sznajdin-MF-raw}
\end{equation}
A similar line of reasoning for the outflow Sznajd model leads to the system

\begin{equation}
\left\{
\begin{aligned}
& \dot{\eta}_{\sigma} = -p \eta_{\sigma} + (1-p)(\eta_{\sigma} + \nu_{\sigma})^2 \sum_{\sigma'\neq \sigma} \nu_{\sigma'} \\
& \dot{\nu}_{\sigma} = p\eta_{\sigma} - (1-p)\nu_{\sigma}\sum_{\sigma'\neq\sigma} (\eta_{\sigma'} + \nu_{\sigma'})^2
\end{aligned}
\right.
\label{eq:sznajdout-MF-raw}
\end{equation}
and equations (\ref{eq:sznajdin-MF-raw}) and (\ref{eq:sznajdout-MF-raw}) can be put in the same form

\begin{equation}
\left\{
\begin{aligned}
& \dot{\eta}_{\sigma} = -\lambda \eta_{\sigma} + (\eta_{\sigma} + \nu_{\sigma})^2 \sum_{\sigma'\neq \sigma} \nu_{\sigma'} \\
& \dot{\nu}_{\sigma} = \lambda\eta_{\sigma} - \nu_{\sigma}\sum_{\sigma'\neq\sigma} (\eta_{\sigma'} + \nu_{\sigma'})^2
\end{aligned}
\right.
\label{eq:sznajd-MF}
\end{equation}
if we change the timescale of the outflow model, such that

\begin{equation}
    \lambda = p_{\mathrm{in}} = \frac{p_{\mathrm{out}}}{1 - p_{\mathrm{out}}}\quad\quad\quad\quad t_{\mathrm{out}} = t_{\mathrm{in}}(1-p_{\mathrm{out}})
\end{equation}
where $p_{\mathrm{in}}$ and $p_{\mathrm{out}}$ are the parameter $p$ in the inflow and outflow models respectively.
Since $\lambda$ can be thought of as the rate at which latent agents turn active, it makes sense to think of $\nicefrac{1}{\lambda}$ as the intensity of the latency. This shows that the main difference between the inflow and outflow versions of the Sznajd model is that we can make the latency arbitrarily low ($\lambda \rightarrow \infty$) in the outflow version, while $\lambda \leq 1$ for the inflow version. Besides that the evolution will be slower in the outflow version because of the change in timescales.

A detailed study of the system (\ref{eq:sznajd-MF}) can be found in appendix \ref{ap:sznajd}. Since the differential equations clearly cannot be solved analytically, we approach the problem qualitatively using linear stability analysis. In order to do that, we must first find the stationary points (also known as fixed points) of (\ref{eq:sznajd-MF}), that is the points $(\eta_1, \nu_1, \ldots, \eta_M,\nu_M)$ where all derivatives $\dot{\eta}_{\sigma}$ and $\dot{\nu}_{\sigma}$ are 0. After that, we linearize the equations around each of these fixed points and examine if small perturbations drive the system closer (stable fixed points) or away from these points (unstable fixed points). The idea being that in the long time limit the system will never end up in an unstable fixed point and barring more complicated situations (like periodicity or chaos) it will end up in one of the stable fixed points.

For the Sznajd model with latency, the fixed points can all be found analytically (see appendix \ref{subap:existence-sznajd}) and are of 2 types

\subsubsection{Asymmetric Coexistences}

In these fixed points, we have one opinion with a proportion of latent agents $\eta_1$ and a proportion of active agents $\nu_1$, while other $n\geq 1$ opinions have a proportion of latent and active agents given by $\eta_2$ and $\nu_2$ respectively. All remaining $M-(n+1)$ opinions are absent. $\eta_{1(2)}$, $\nu_{1(2)}$ are given by

\begin{equation}
\left\{
\begin{aligned}
&\zeta = \frac{1-2\lambda n \pm \sqrt{1 - 4\lambda(2n-1)}}{2\lambda}\\
& \nu_{1} = \frac{\lambda\zeta(\zeta + n)}{\lambda(\zeta + n)^2 + n} \\
& \nu_{2} = \frac{\lambda(\zeta + n)}{\zeta(\lambda(\zeta + n)^2 + n)}\\
&\eta_{1} = \frac{\zeta}{\zeta + n} - \nu_1 \\
&\eta_{2} = \frac{1}{\zeta + n} - \nu_2
\end{aligned}
\right.
\end{equation}
Furthermore, the fixed points for a particular value of $n$ exist iff $\lambda \leq \nicefrac{1}{(8n-4)}$. Studying how these fixed points behave under small perturbations, we find that they are always unstable (see appendix \ref{subap:stability-sznajd}).

\subsubsection{Symmetric Coexistences}

In these fixed points, we have $n$ opinions with a proportion of latent agents $\eta$ and a proportion of active agents $\nu$, while all remaining $M-n$ opinions are absent. $\eta$, $\nu$ are given by

\begin{equation}
\left\{
\begin{aligned}
&\nu = \frac{\lambda n}{\lambda n^2 + n - 1}\\
&\eta = \frac{1}{n} - \nu\\
\end{aligned}
\right.
\end{equation}
These fixed points always exist for all $n$ between 1 and $M$, however, what we find when we consider small perturbations is that these points are always stable for $n=1$, they are stable iff $\lambda< \nicefrac{1}{4}$ when $n=2$ and they are always unstable when $n>2$ (again, see appendix \ref{subap:stability-sznajd}).

The fixed points with $n=1$ are consensus states (and we have one for each of the $M$ opinions), while the fixed points with $n=2$ are a symmetrical coexistence between exactly two opinions (we have one for each pair of opinions possible).

Together with the results of the asymmetric coexistences, this explains what we saw in the simulations for the Barabási-Albert network. Any situation with more than 2 opinions present is ultimately unstable and unless latency is high enough, the long time limit is always a consensus state.

\begin{figure}[htbp]
    \centering
    \includegraphics[width=0.35\textwidth]{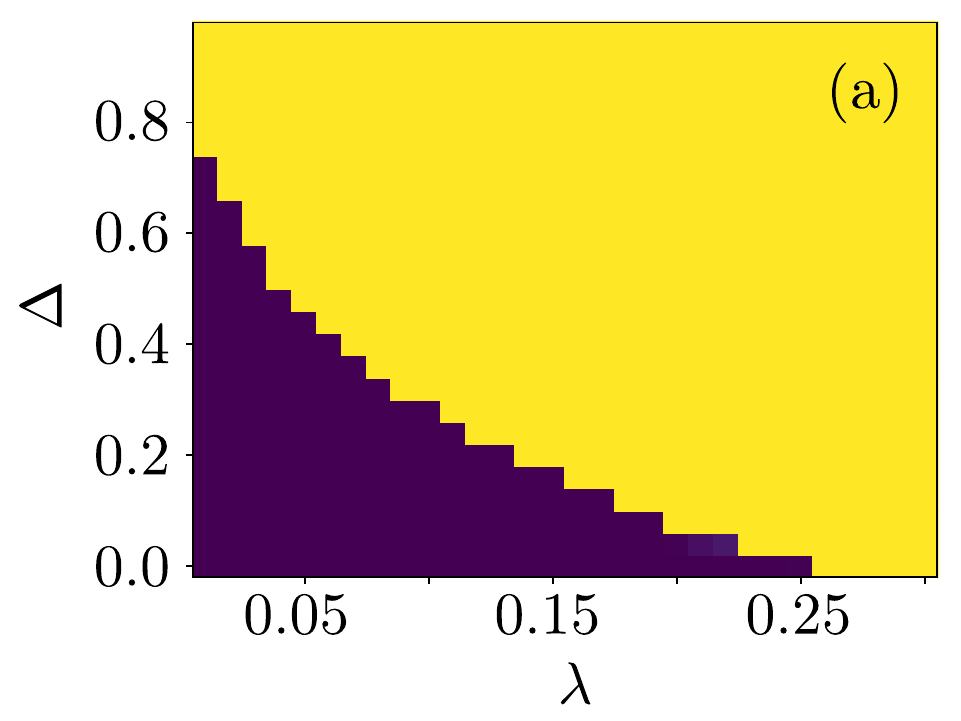} \includegraphics[width=0.35\textwidth]{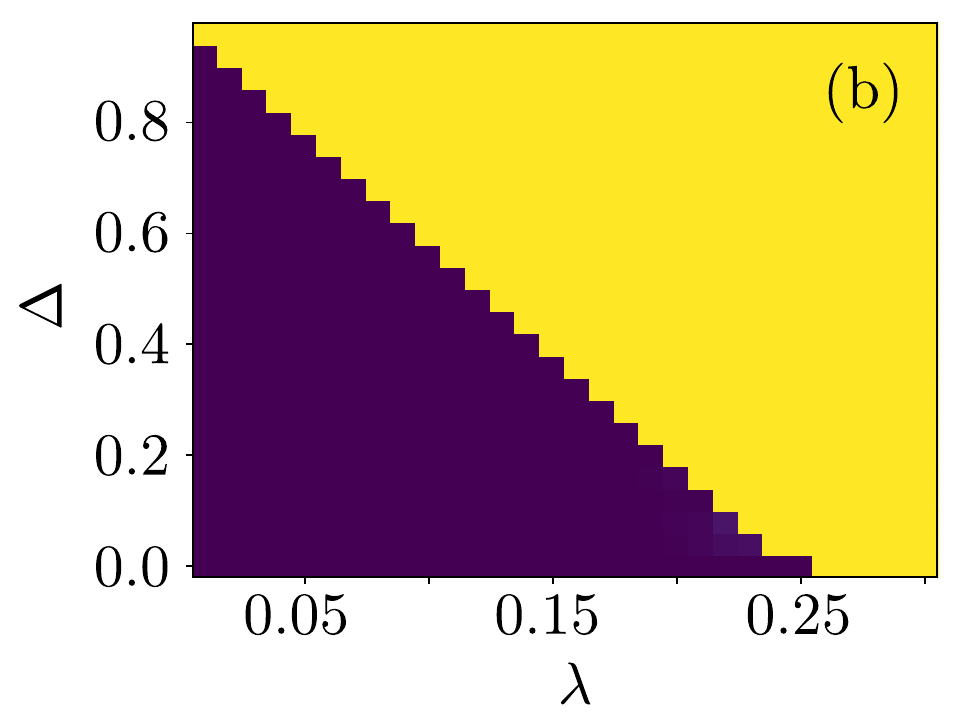}
    \caption{Parameters for which the mean field equations of the Sznajd model converge to consensus (yellow) or to a symmetric coexistence (purple) as a function of $\lambda$ and of the asymmetry $\Delta = |\theta_{1\circ} - \theta_{2\circ}|$, where $\theta_{1(2)\circ}$ is the starting proportion of agents with opinion 1 (2). (a) With the initial condition such that all agents are active, matching what was done in the simulations. (b) With the initial condition such that all agents are latent.}
    \label{fig:mf-bacia}
\end{figure}

Integrating the mean field equations (\ref{eq:sznajd-MF}) numerically allows us to obtain more information about when the model evolves towards consensus or coexistence. In figure \ref{fig:mf-bacia} we can see the values of $p$ leading to each type of final state depending on how asymmetric the initial condition is (compare with figure \ref{fig:sims-bacia} where the same thing is studied in the simulations).

\subsection{Voter Model}

Using a similar reasoning for the voter model, we get the following system of differential equations

\begin{equation}
    \left\{
    \begin{aligned}
        & \dot{\eta}_{\sigma} = -p \eta_{\sigma} + (\eta_{\sigma} + \nu_{\sigma})\sum_{\sigma'\neq \sigma} \nu_{\sigma'} \\
        & \dot{\nu}_{\sigma} = p \eta_{\sigma} - \nu_{\sigma} \sum_{\sigma'\neq \sigma} (\eta_{\sigma'} + \nu_{\sigma'})
    \end{aligned}
    \right.
    \label{eq:voter-MF}
\end{equation}
The detailed study of the system (\ref{eq:voter-MF}) can be found in the appendix \ref{ap:voter}. Like with the Sznajd model, we did a linear stability analysis. All fixed points for the voter model are such that we have $n$ opinions with a proportion of latent agents $\eta$ and a proportion of active agents $\nu$, while all remaining $M-n$ opinions are absent. $\eta$, $\nu$ are given by

\begin{equation}
\left\{
\begin{aligned}
& \nu = \frac{\lambda}{\lambda n + n - 1}\\
& \eta = \frac{1}{n} - \nu
\end{aligned}
\right.
\end{equation}
These fixed points always exist for all $n$ between 1 and $M$. Considering small perturbations, we see that these fixed points are unstable for $n<M$ and stable for $n=M$ (see appendix \ref{ap:voter}).

This explains what we observed in the simulations with the Barabási-Albert network. The only stable situation is a symmetric coexistence of all opinions. As a consequence, even if we start from extremely asymmetric initial conditions, none of the opinions disappear, and the model evolves towards a symmetric coexistence.

\section{Conclusions}
\label{sec:conc}

In the simulations with Barabási-Albert networks and in the mean field treatment developed, we were able to observe that the Sznajd model with latency allows for the existence of stable opinion splits involving 2 opinions, but states with more than 2 opinions are always unstable. This is in contrast with what we observe for the voter model, where the only stable equilibrium for the model with $M$ opinions is the symmetric coexistence of all the $M$ opinions. This allows us to draw the following conclusions

\begin{itemize}
    \item Latency and noise seem to behave similarly when applied to a binary opinion propagation model and can both be thought of as additions that create a tendency towards coexistence of opinions instead of consensus. However the example of the Sznajd model with latency shows that there are fundamental differences between these two model ingredients. This can be seem if we compare our results with the behaviour of the Sznajd model with noise \cite{Maycon} where we either end up in a state with a clear majority opinion or in a coexistence of all opinions.
    \item The Sznajd model with latency does a better job at justifying the use of binary opinion models compared with the voter model, since if situations with more than 2 options are unstable, then they can be safely ignored.
    \item Based on these results, we speculate that a mechanism similar to latency might result in the emergence of political spectra, when extended to continuous opinion models.
\end{itemize}

\begin{acknowledgments}
The authors would like to thank CAPES for the financial support and the Multi-user Computing Center
(CCM/Propes) of UFABC for the computational resources used for the simulations.
\end{acknowledgments}

\bibliography{sociophysics}

\pagebreak
\widetext
\appendix
\onecolumngrid
\setcounter{equation}{0}
\setcounter{figure}{0}
\setcounter{table}{0}

\section{Detailed Calculations}
\label{ap:sznajd}

The equations for the Sznajd model with latency in the mean field approximation are

\begin{equation}
\left\{
\begin{aligned}
& \dot{\eta}_{\sigma} = -\lambda \eta_{\sigma} + (\eta_{\sigma} + \nu_{\sigma})^2 \sum_{\sigma'\neq \sigma} \nu_{\sigma'} \\
& \dot{\nu}_{\sigma} = \lambda\eta_{\sigma} - \nu_{\sigma}\sum_{\sigma'\neq\sigma} (\eta_{\sigma'} + \nu_{\sigma'})^2
\end{aligned}
\right.
\end{equation}
Where $\eta_{\sigma}$ is the proportion of sites in the network that are not susceptible and have opinion $\sigma$, while $\nu_{\sigma}$ is the proportion of sites in the network that are susceptible and have opinion $\sigma$. Defining

\begin{equation}
\theta_{\sigma} = \eta_{\sigma} + \nu_{\sigma}
\end{equation}
\begin{equation}
\nu = \sum_{\sigma}\nu_{\sigma}\quad\mbox{and}\quad Q^2 = \sum_{\sigma} \theta_{\sigma}^2.
\end{equation}
Then $\sum_{\sigma} \theta_{\sigma} = 1$  and we can rewrite the equations of motion as

\begin{equation}
\left\{
\begin{aligned}
& \dot{\theta}_{\sigma} = \theta_{\sigma}^2\nu - Q^2\nu_{\sigma} \\
& \dot{\nu}_{\sigma} = \lambda\theta_{\sigma} - \nu_{\sigma}(Q^2 + \lambda - \theta_{\sigma}^2)
\end{aligned}
\right.
\label{eq:MF-eq-theta}
\end{equation}

\subsection{Existence of Fixed Points}
\label{subap:existence-sznajd}

In the fixed points we have $\dot{\theta}_{\sigma} = 0$ and $\dot{\nu}_{\sigma}=0$. From $\dot{\theta}_{\sigma} = 0$ it follows that in a fixed point

\begin{equation}
\theta_{\sigma}^2\nu - Q^2\nu_{\sigma} = 0 \Leftrightarrow \frac{\nu_{\sigma}}{\nu} = \frac{\theta_{\sigma}^2}{Q^2} \equiv \xi_{\sigma}^2 \Rightarrow
\end{equation}
\begin{equation}
\nu_{\sigma} = \nu \xi_{\sigma}^2\quad\mbox{and}\quad \theta_{\sigma} = Q\xi_{\sigma}
\end{equation}
Substituting in $\dot{\nu}_{\sigma}=0$ yields

\begin{equation}
\lambda Q\xi_{\sigma} - \nu \xi_{\sigma}^2 (Q^2 + \lambda -Q^2\xi^2_{\sigma}) = 0
\end{equation}
A trivial solution is $\xi_{\sigma} = 0$, meaning the opinion is extinct. Removing this case we arrive at

\begin{equation}
\nu Q^2\xi_{\sigma}^3 - \nu(Q^2 + \lambda) \xi_{\sigma} +\lambda Q = 0
\end{equation}
Multiplying by $\nicefrac{Q}{\nu}$ on both sides we can rewrite this equation as an equation for $\theta_{\sigma}$:

\begin{equation}
\theta_{\sigma}^3 - (Q^2 + \lambda) \theta_{\sigma} + \frac{\lambda Q^2}{\nu} = 0
\end{equation}
Using Girard-Vieta relations, it follows that this equation has an odd number of negative real roots (which can't be used as solutions) and all the roots add up to 0. This means that we must have exactly one negative root and the other 2 roots have positive real part. The important thing is that all coordinates $\theta_{\sigma}$ can have at most 2 values $\theta_{\pm}$. In other words, if $Q$ and $\nu$ are assumed as parameters, $\theta_{\sigma} = \theta_{\pm}$ gives a solution. Since $Q$ and $\nu$ actually depend on the solution itself, we need to go back to the fixed point equations and find a self-consistent solution. Let $n_{\pm}$ be the number of opinions with proportion $\theta_{\pm}$. It follows that

\begin{equation}
Q^2 = n_{-}\theta_{-}^2 + n_{+}\theta_{+}^2\quad\mbox{and}\quad \nu = n_{-}\nu_{-} + n_{+}\nu_{+}
\end{equation}
We also impose normalization by adding the constraint

\begin{equation}
n_{-}\theta_{-} + n_{+}\theta_{+} = 1
\end{equation}
It follows that the fixed point equations are reduced to 4 equations: $\dot{\theta}_{\pm} = 0$ and $\dot{\nu}_{\pm} = 0$.

\begin{equation}
\dot{\theta}_{\pm} = 0 \Leftrightarrow \theta_{\pm}^2(n_{-}\nu_{-} + n_{+}\nu_{+}) = \nu_{\pm}(n_{-}\theta_{-}^2 + n_{+}\theta_{+}^2)
\end{equation}
The solution for this is

\begin{equation}
\nu_{-} = \zeta^2\nu_{+} \quad\mbox{and}\quad \theta_{-} = \zeta\theta_{+}
\end{equation}
for $\zeta > 0$. Imposing normalization allows us to write $\zeta$ and $\theta_{+}$ in terms of each other:

\begin{equation}
n_{-}\theta_{-} + n_{+}\theta_{+} = 1 \Rightarrow \theta_{+}(n_{-}\zeta + n_{+}) = 1 \Rightarrow \frac{1}{\theta_{+}} = n_{-}\zeta + n_{+},
\end{equation}
\begin{equation}
\theta_{+} = \frac{1}{n_{-}\zeta + n_{+}}\quad\mbox{and}\quad \theta_{-} = \frac{\zeta}{n_{-}\zeta + n_{+}}
\end{equation}
Next

\begin{equation}
\dot{\nu}_{\pm} = 0 \Leftrightarrow \lambda \theta_{\pm} = \nu_{\pm}(n_{-}\theta_{-}^2 + n_{+}\theta_{+}^2 + \lambda -\theta_{\pm}^2)
\end{equation}
that yields

\begin{equation}
\nu_{-} = \frac{\nicefrac{\lambda\zeta}{\theta_{+}}}{\nicefrac{\lambda}{\theta_{+}^2} + (n_{-}-1)\zeta^2 + n_{+}} = 
\frac{\lambda\zeta(n_{-}\zeta + n_{+})}{\lambda(n_{-}\zeta + n_{+})^2 + (n_{-}-1)\zeta^2 + n_{+}}
\end{equation}
and

\begin{equation}
\nu_{+} = \frac{\nicefrac{\lambda}{\theta_{+}}}{\nicefrac{\lambda}{\theta_{+}^2} + n_{-}\zeta^2 + n_{+} -1} = 
\frac{\lambda(n_{-}\zeta + n_{+})}{\lambda(n_{-}\zeta + n_{+})^2 + n_{-}\zeta^2 + n_{+} -1}
\end{equation}
Finally $\nu_{-} = \zeta^2\nu_{+}$ gives as an equation for $\zeta$:

\begin{equation}
\frac{\lambda\zeta(n_{-}\zeta + n_{+})}{\lambda(n_{-}\zeta + n_{+})^2 + (n_{-}-1)\zeta^2 + n_{+}} = \frac{\lambda\zeta^2(n_{-}\zeta + n_{+})}{\lambda(n_{-}\zeta + n_{+})^2 + n_{-}\zeta^2 + n_{+} -1} \Rightarrow
\end{equation}
\begin{equation}
\lambda(n_{-}\zeta + n_{+})^2 + n_{-}\zeta^2 + n_{+} -1 = \zeta(\lambda(n_{-}\zeta + n_{+})^2 + (n_{-}-1)\zeta^2 + n_{+}) \Rightarrow
\end{equation}
\begin{equation}
(\zeta-1)(\lambda(n_{-}\zeta + n_{+})^2 + n_{-}\zeta^2 + n_{+} - \zeta^2 - \zeta - 1) = 0
\end{equation}
This gives a solution $\zeta = 1$ that corresponds to a symmetric coexistence ($\theta_{+} = \theta_{-}, \nu_{+} = \nu_{-}$). Removing this solution we have

\begin{equation}
\lambda(n_{-}\zeta + n_{+})^2 + n_{-}\zeta^2 + n_{+} - \zeta^2 - \zeta - 1 = 0 \Rightarrow
\end{equation}
\begin{equation}
(\lambda n_{-}^2 + n_{-} - 1)\zeta^2 + (2\lambda n_{-}n_{+} - 1)\zeta + (\lambda n_{+}^2 + n_{+} - 1) = 0
\end{equation}
whose solutions display the symmetry $n_{-} \leftrightarrow n_{+} \Leftrightarrow \zeta \leftrightarrow \nicefrac{1}{\zeta}$, which reflects the way $\zeta$ is defined. Looking at the discriminant of that equation as a polynomial for $\zeta$ we have

\begin{equation}
\Delta = (2\lambda n_{-}n_{+} - 1)^2 - 4(\lambda n_{-}^2 + n_{-} - 1)(\lambda n_{+}^2 + n_{+} - 1) = 
\end{equation}
\begin{equation}
= 1 - 4(n_{-} - 1)(n_{+} - 1) - 4\lambda(n_{-}n_{+} + (n_{-}-1)n_{+}^2 + n_{-}^2(n_{+}-1))
\end{equation}
Hence $\Delta \geq 0$ only if $n_{-} = 1$ or $n_{+} = 1$. For $n_{-}=1$ and $n_{+}=n$, the equation for $\zeta$ becomes

\begin{equation}
\lambda\zeta^2 + (2\lambda n - 1)\zeta + (\lambda n^2 + n - 1) = 0\quad\mbox{and}\quad \Delta = 1 - 4\lambda(2n-1)
\end{equation}
So the corresponding asymmetric solution is

\begin{equation}
n_{-} = 1\,\,\mbox{and}\,\, n_{+}=n \Rightarrow
\left\{
\begin{aligned}
&\zeta = \frac{1-2\lambda n \pm \sqrt{1 - 4\lambda(2n-1)}}{2\lambda}\\
&\theta_{+} = \frac{1}{\zeta + n}\\
&\theta_{-} = \frac{\zeta}{\zeta + n}\\
&\nu_{+} = \frac{\lambda(\zeta + n)}{\zeta(\lambda(\zeta + n)^2 + n)}\\
&\nu_{-} = \frac{\lambda\zeta(\zeta + n)}{\lambda(\zeta + n)^2 + n}\\
\end{aligned}
\right.
\end{equation}
Where the $(+)$ solution occurs in $n$ opinions and the $(-)$ solution occurs in only one. The $n_{-} \leftrightarrow n_{+} \Leftrightarrow \zeta \leftrightarrow \nicefrac{1}{\zeta}$ symmetry implies that the solution for $n_{+}=1$ and $n_{-}=n$ is equivalent to this one ($\zeta$ is differrent, but the $(\theta, \nu)$ coordinates are the same). Finally this solution exists for

\begin{equation}
\lambda \leq \frac{1}{8n-4}\quad\mbox{and}\quad \lambda = \frac{1}{8n-4} \Rightarrow \zeta = 3n-2
\end{equation}
so for $n=1$ (which is the case where we have 2 different opinions), the biffurcation that creates/annihilates this solution coincides with the symmetric coesxistence of these opinions.

The symmetric solution ($\zeta = 1$) is easy to recover

\begin{equation}
\left\{
\begin{aligned}
&\theta_{\sigma} = \frac{1}{n+1}\\
&\nu_{\sigma} = \frac{\lambda(n+1)}{\lambda(n+1)^2 + n}\\
\end{aligned}
\right.
\end{equation}
for the same number of opinions ($n+1$) used in the asymmetric solution.

Since the equations separate, we can add to any of these solutions any number of extinct opinions ($\theta_{\sigma} = \nu_{\sigma} = 0$)

\subsection{Stability}
\label{subap:stability-sznajd}

For the jacobian of the time evolution, recall that

\begin{equation}
\frac{\partial Q^2}{\partial \theta_{\sigma}} = 2\theta_{\sigma}, \quad \frac{\partial Q^2}{\partial \nu_{\sigma}} = 0, \quad \frac{\partial \nu}{\partial \theta_{\sigma}} = 0, \quad \frac{\partial \nu}{\partial \nu_{\sigma}} = 1
\end{equation}
So

\begin{equation}
\left\{
\begin{aligned}
& \dot{\theta}_{\sigma} = \theta_{\sigma}^2\nu - Q^2\nu_{\sigma} \\
& \dot{\nu}_{\sigma} = \lambda\theta_{\sigma} - \nu_{\sigma}(Q^2 + \lambda - \theta_{\sigma}^2)
\end{aligned}
\right.
\end{equation}
implies

\begin{equation}
\left\{
\begin{aligned}
&\frac{\partial \dot{\theta}_{\sigma}}{\partial \theta_{\sigma'}} = 2\theta_{\sigma}\nu\delta_{\sigma,\sigma'} - 2\theta_{\sigma'}\nu_{\sigma}\\
&\frac{\partial \dot{\theta}_{\sigma}}{\partial \nu_{\sigma'}} = - Q^2\delta_{\sigma,\sigma'} + \theta_{\sigma}^2\\
&\frac{\partial \dot{\nu}_{\sigma}}{\partial \theta_{\sigma'}} = \delta_{\sigma,\sigma'}(\lambda + 2\theta_{\sigma}\nu_{\sigma}) - 2\theta_{\sigma'}\nu_{\sigma}\\
&\frac{\partial \dot{\nu}_{\sigma}}{\partial \nu_{\sigma'}} = \delta_{\sigma,\sigma'}( \theta_{\sigma}^2 -Q^2 - \lambda)\\
\end{aligned}\right.
\end{equation}
Lets break up the opinions in our fixed point in 3 sets:

\begin{itemize}
\item $\Omega$: With the extinct opinions
\item $\Delta_1$: With the opinion that appears only once (in case we are studying an asymmetric fixed point)
\item $\Delta_n$: With the remaining opinions
\end{itemize}

It follows that $\mathcal{J}_{\Omega, \Delta_1}$ and $\mathcal{J}_{\Omega, \Delta_n}$ are null, so the eigenvalues of $\mathcal{J}_{\Omega,\Omega}$ separate. Since

\begin{equation}
\mathcal{J}_{\Omega,\Omega} =
\begin{bmatrix}
0 & -Q^2 \mathds{I}\\
\lambda \mathds{I} & -(Q^2 + \lambda)\mathds{I}
\end{bmatrix}
\label{eq:jac-sz-omega}
\end{equation}
this matrix has as eigenvalues $-\lambda$ and $-Q^2$, both with multiplicity $|\Omega|$, meaning that the corresponding manifold is attractive ($2|\Omega|$ attractive directions).

For the remaning eigenvalues, note that if we organize $\mathcal{J}_{\Delta, \Delta}$ as

\begin{equation}
\begin{bmatrix}
(\theta_n, \theta_n) & (\theta_n, \nu_n) & (\theta_n, \theta_1) & (\theta_n, \nu_1) \\
(\nu_n, \theta_n) & (\nu_n, \nu_n) & (\nu_n, \theta_1) & (\nu_n, \nu_1) \\
(\theta_1, \theta_n) & (\theta_1, \nu_n) & (\theta_1, \theta_1) & (\theta_1, \nu_1) \\
(\nu_1, \theta_n) & (\nu_1, \nu_n) & (\nu_1, \theta_1) & (\nu_1, \nu_1)
\end{bmatrix}
\end{equation}
and define $\vec{1} = (1,\ldots ,1)\in \mathds{R}^n$, $\mathds{E} = \vec{1}\otimes \vec{1}$, then in the case of an asymmetric fixed point

\begin{equation}
\mathcal{J}_{\Delta, \Delta} = \begin{bmatrix}
A\mathds{I} + \alpha \mathds{E} & B\mathds{I} + \beta \mathds{E} & a\vec{1} & b\vec{1} \\
C\mathds{I} + \gamma \mathds{E} & D\mathds{I} + \delta \mathds{E} & c\vec{1} & d\vec{1} \\
e\vec{1} & f\vec{1} & x & y \\
g\vec{1} & h\vec{1} & z & w
\end{bmatrix}
\label{eq:jac-sz-delta}
\end{equation}
which implies that the remaining eigenvalues are the eigenvalues of

\begin{equation}
\begin{bmatrix}
A & B \\
C & D
\end{bmatrix}
\label{eq:subjac1}
\end{equation}
with multiplicity $n-1$, together with the eigenvalues of

\begin{equation}
\begin{bmatrix}
A + \alpha n & B + \beta n & a n & b n \\
C + \gamma n & D + \delta n & c n & d n \\
e & f & x & y \\
g & h & z & w
\end{bmatrix}
\label{eq:subjac2}
\end{equation}
with multiplicity 1 (see appendix \ref{ap:extra}). Whereas in the case of a symmetric fixed point we have

\begin{equation}
\mathcal{J}_{\Delta, \Delta} = \begin{bmatrix}
A\mathds{I} + \alpha \mathds{E} & B\mathds{I} + \beta \mathds{E} \\
C\mathds{I} + \gamma \mathds{E} & D\mathds{I} + \delta \mathds{E} \\
\end{bmatrix}
\label{eq:jac-sz-delta-simple}
\end{equation}
and the eigenvalues are the ones of

\begin{equation}
\begin{bmatrix}
A & B \\
C & D
\end{bmatrix}
\quad\mbox{and}\quad
\begin{bmatrix}
A + \alpha n & B + \beta n \\
C + \gamma n & D + \delta n
\end{bmatrix}
\end{equation}

\subsubsection{Symmetric Fixed Points}

For the symmetric fixed points with $M$ opinions we have

\begin{equation}
\mathcal{J}_{\Delta, \Delta} = \begin{bmatrix}
A\mathds{I} + \alpha \mathds{E} & B\mathds{I} + \beta \mathds{E} \\
C\mathds{I} + \gamma \mathds{E} & D\mathds{I} + \delta \mathds{E} \\
\end{bmatrix}
\end{equation}
If $M$ is the number of opinions, we have

\begin{equation}
\theta_{\sigma} = \frac{1}{M},\quad\nu_{\sigma} = \frac{\lambda M}{\lambda M^2 + M - 1},\quad\nu = \frac{\lambda M^2}{\lambda M^2 + M - 1},\quad Q^2 = \frac{1}{M} \Rightarrow
\end{equation}
\begin{equation}
\left\{
\begin{aligned}
& A = 2\nu\theta_{\sigma} = \frac{2\lambda M}{\lambda M^2 + M - 1}\\
& B = -Q^2 = \frac{-1}{M} \\
& C = \lambda + 2\theta_{\sigma}\nu_{\sigma} = \lambda + \frac{2\lambda}{\lambda M^2 + M - 1} \\
& D = \theta_{\sigma}^2 - Q^2 - \lambda = \frac{1}{M^2} - \frac{1}{M} - \lambda
\end{aligned}
\right.
\end{equation}
It follows that

\begin{equation}
\mathrm{det}
\begin{bmatrix}
A & B \\
C & D
\end{bmatrix}
= \frac{\lambda(3-M-\lambda M^2)}{M(\lambda M^2 + M - 1)}
\end{equation}
and
\begin{equation}
\mathrm{Tr}
\begin{bmatrix}
A & B \\
C & D
\end{bmatrix}
= \frac{M(2-M) - (\lambda M^2 - 1)^2}{M^2(\lambda M^2 + M - 1)}
\end{equation}
A necessary condition for the stability of this fixed point is $\mathrm{Tr} < 0$ and $\mathrm{det} > 0$. If $M\geq 3$ we have $\mathrm{det} < 0$ indicating the solutions are unstable. Since this matrix is irrelevant for the stability of the $M=1$ solution (the eigenvalues coming from it appear $M-1$ times), we only need to look at the $M=2$ case:

\begin{equation}
\mathrm{Tr} = \frac{- (4\lambda - 1)^2}{4(4\lambda + 1)} < 0\quad \mbox{iff} \quad\lambda \neq \frac{1}{4}
\end{equation}
\begin{equation}
\mathrm{det} = \frac{1-4\lambda}{\lambda} > 0\quad \mbox{iff} \quad\lambda < \frac{1}{4}
\end{equation}
so the solution is unstable for $\lambda > \nicefrac{1}{4}$ and the stability for $\lambda < \nicefrac{1}{4}$ depends on the eigenvalues of the second matrix (note that $M=2$, $\lambda = \nicefrac{1}{4}$ is a bifurcation leading to the appearance of asymmetric solutions at the position of the symmetric solution)

The second matrix

\begin{equation}
\begin{bmatrix}
A + \alpha M & B + \beta M \\
C + \gamma M & D + \delta M
\end{bmatrix}
\end{equation}
must have 0 as an eigenvalue, corresponding to the embedding artifact (more details in appendix \ref{ap:extra}), so the remaining relevant eigenvalue is just its trace. We also only need to study the cases $M=1,2$ as $M\geq 3$ has already been determined to be unstable.

We have

\begin{equation}
\alpha = -2\theta_{\sigma'}\nu_{\sigma} = \frac{2\lambda}{\lambda M^2 + M - 1} \Rightarrow A + \alpha M = 0 \quad\mbox{and}\quad \delta = 0\Rightarrow
\end{equation}
\begin{equation}
\mathrm{Tr}
\begin{bmatrix}
A + \alpha M & B + \beta M \\
C + \gamma M & D + \delta M
\end{bmatrix}
= D = \frac{1}{M^2} - \frac{1}{M} - \lambda < 0 \quad\mbox{if}\quad \lambda > 0 \,\,\mbox{or}\,\, M>1
\end{equation}
So fixed points with $M = 1$ are stable for $\lambda > 0$, symmetric fixed points with $M=2$ are stable iff $\lambda < \nicefrac{1}{4}$ and all other symmetric fixed points are unstable.

\subsubsection{Asymmetric Fixed Points}

For the asymmetric fixed points with $n+1$ opinions we have

\begin{equation}
\left\{
\begin{aligned}
& \theta_n = \frac{1}{n + \zeta} \\
& \nu_n = \frac{\lambda(n+\zeta)}{\zeta(\lambda(n+\zeta)^2 + n)} \\
& \nu = (n+\zeta^2) \nu_n \\
& Q^2 = (n+\zeta^2) \theta_n^2
\end{aligned}
\right.
\end{equation}
So that using these and the quadratic equation that defines $\zeta$ we have

\begin{equation}
\mathrm{det}
\begin{bmatrix}
A & B \\
C & D
\end{bmatrix}
= \frac{\nu_n(n+\zeta^2)}{\lambda^2(n+\zeta)^3}(\zeta(\lambda^2(n-1) - \lambda) + \lambda^2(n^2 + n + 2) + \lambda(n-1))
\end{equation}
and the sign of this determinant is just the sign of

\begin{equation}
\zeta(\lambda^2(n-1) - \lambda) + \lambda^2(n^2 + n + 2) + \lambda(n-1)
\end{equation}
Substituting the 2 solutions $\zeta_{\pm}$ for the corresponding quadratic equation and multiplying by 2 yields

\begin{equation}
(1-2\lambda n \pm \sqrt{1-4\lambda(2n-1)})
 (\lambda n-\lambda - 1) + 2\lambda^2(n^2 + n + 2) + 2\lambda(n-1) =
\end{equation}
\begin{equation}
= 4\lambda^2(n+1) + 5\lambda n - 3\lambda - 1 \pm (\lambda n-\lambda - 1)\sqrt{1-4\lambda(2n-1)} \equiv \xi_{n\pm}(\lambda)
\end{equation}
To test the sign of $\xi_{n\pm}(\lambda)$, note that the equation $\xi_{n\pm}(\lambda) = 0$ implies that

\begin{equation}
16\lambda^4(n+1)^2 + 4\lambda^3(2n^3 + 5n^2 + 8n - 7) + 4\lambda^2(2n^2 - 3n - 2) = 0 \Leftrightarrow
\end{equation}
\begin{equation}
\lambda^2(4\lambda + 2n + 1)((n+1)^2\lambda + n - 2) = 0 \Rightarrow \lambda = 0\quad\mbox{or}\quad\lambda = \frac{-(1+2n)}{4}\quad\mbox{or}\quad \lambda = \frac{2-n}{(n+1)^2}
\end{equation}
The eigenvalues of the matrix 

\begin{equation}
\begin{bmatrix}
A & B \\
C & D
\end{bmatrix}
\end{equation}
are relevant only for points with $n > 1$, meaning that $\xi_{n\pm}(\lambda)$ doesn't change signs as $\lambda$ changes (as long as it remains in the $\lambda > 0$ range). So expanding for small $\lambda$ we get

\begin{equation}
\xi_{n+}(\lambda) \simeq -2 \quad\forall\,\, n > 1
\end{equation}
\begin{equation}
\xi_{n-}(\lambda) \simeq -2\lambda^2(n - 2)(2n + 1) < 0 \quad\forall\,\, n > 2
\end{equation}
\begin{equation}
\xi_{2-}(\lambda) \simeq -90\lambda^3
\end{equation}
which shows that

\begin{equation}
\mathrm{det}
\begin{bmatrix}
A & B \\
C & D
\end{bmatrix} < 0
\end{equation}
for all asymmetric fixed points with $n>1$, meaning they are all unstable. In the case $n=1$, $\mathcal{J}_{\Delta,\Delta}$ is
\begin{equation}
\mathcal{J}_{\Delta,\Delta}=
\begin{bmatrix}
2\theta_+\nu_- & -2\theta_-\nu_+ & -\theta^2_- & \theta^2_+ \\
-2\theta_+\nu_- & 2\theta_-\nu_+ & \theta^2_- & -\theta^2_+ \\
\lambda & -2\theta_-\nu_+ & -(\lambda + \theta^2_-) & 0 \\
-2\theta_+\nu_- & \lambda & 0 & -(\lambda + \theta^2_+)
\end{bmatrix}
\end{equation}
The embedding artifact comes from the left eigenvector $[1, 1, 0, 0]$, so by duality it can be removed, resulting in the matrix

\begin{equation}
\widetilde{\mathcal{J}} = 
\begin{bmatrix}
2\theta_+\nu_- +2\theta_-\nu_+ & -\theta^2_- & \theta^2_+ \\
\lambda +2\theta_-\nu_+ & -(\lambda + \theta^2_-) & 0 \\
-\lambda - 2\theta_+\nu_-  & 0 & -(\lambda + \theta^2_+)
\end{bmatrix}
\end{equation}
From this, substituting the fixed point and using the quadratic equation that defines $\zeta$ we get

\begin{equation}
2\mathrm{det}(\widetilde{\mathcal{J}})(\zeta + 1)^4(\lambda(\zeta + 1)^2 + 1)\lambda^3 = 
2\lambda\zeta(1 - 7\lambda + 13\lambda^2 -4\lambda^3) -8\lambda^4 + 10\lambda^3 - 2\lambda^2
\end{equation}
that has the same sign as $\mathrm{det}(\widetilde{\mathcal{J}})$. Substituting the solution for $\zeta$ yields
\begin{equation}
1 - 9\lambda + 25\lambda^2 -20\lambda^3 \pm (1 - 7\lambda + 13\lambda^2 -4\lambda^3)\sqrt{1 -4\lambda} \equiv \psi_{\pm}(\lambda)
\end{equation}
Like in our analysis of the $n > 1$ case, we write $\psi_{\pm}(\lambda) = 0$ which leads to
\begin{equation}
(4\lambda-1)^2\lambda^5 = 0
\end{equation}
so the sign of $\psi_{\pm}(\lambda)$ doesn't change in the interval $[0,\nicefrac{1}{4}]$ (which covers all values of $\lambda$ such that this fixed point exists), so making an expansion for small $\lambda$ we have

\begin{equation}
\psi_{+} \simeq 2\quad\mbox{and}\quad \psi_{-} \simeq 2\lambda^5
\end{equation}
so $\det(\widetilde{\mathcal{J}}) > 0$. This means the fixed points must be unstable (if all eigenvalues had negative real part, since the matrix is real, an odd number of them would also need to be real implying their product and hence the determinant would be negative. On the other hand a positive determinant indicates 1 or 3 eigenvalues with positive real part).

Putting everything together, this means all the asymmetric fixed points are unstable for $\lambda > 0$.

\section{Voter Mean Field}
\label{ap:voter}

The equations for the voter model with $M$ opinions and latency in the mean field approximation are

\begin{equation}
\left\{
\begin{aligned}
& \dot{\eta}_{\sigma} = -\lambda \eta_{\sigma} + \sum_{\sigma'\neq \sigma} (\eta_{\sigma} + \nu_{\sigma})\nu_{\sigma'} \\
& \dot{\nu}_{\sigma} = \lambda \eta_{\sigma} - \nu_{\sigma}\sum_{\sigma'\neq \sigma} (\eta_{\sigma'} + \nu_{\sigma'}) 
\end{aligned}
\right.
\label{eq:CM-voter}
\end{equation}
Where $\eta_{\sigma}$ is the proportion of sites in the network that are not susceptible and have opinion $\sigma$, while $\nu_{\sigma}$ is the proportion of sites in the network that are susceptible and have opinion $\sigma$. 
Defining

\begin{equation}
\theta_{\sigma} = \eta_{\sigma} + \nu_{\sigma}\quad\quad S=\sum_{\sigma} \theta_{\sigma} \quad\quad\mbox{and}\quad\nu = \sum_{\sigma}\nu_{\sigma}
\end{equation}
this leads to

\begin{equation}
\left\{
\begin{aligned}
& \dot{\theta}_{\sigma} = \theta_{\sigma}\nu - \nu_{\sigma}S \\
& \dot{\nu}_{\sigma} = \lambda \theta_{\sigma} - \nu_{\sigma}(\lambda + S - \theta_{\sigma}) 
\end{aligned}
\right.
\label{eq:CM-voter2}
\end{equation}
So in the fixed points we must have

\begin{equation}
\nu_{\sigma} = \frac{\theta_{\sigma}\nu}{S}
\end{equation}
and
\begin{equation}
\lambda \theta_{\sigma} S = \theta_{\sigma}\nu(\lambda + S - \theta_{\sigma}).
\end{equation}
Similarly to what happened in the case of the Sznajd model, this implies that $\theta_{\sigma}$ can only have two possible values:

\begin{equation}
\theta_{\sigma} = 0 \quad\mbox{or}\quad \theta_{\sigma} = \frac{\nu\lambda + \nu S -\lambda S}{\nu}
\end{equation}
So if we define $\Omega$ and $\Delta$ as the sets with the extinct opinions and the remaining opinions (respectively), then summing over all $\theta_{\sigma}$ leads to

\begin{equation}
S = \sum_{\sigma} \theta_{\sigma} = \frac{|\Delta|}{\nu}(\nu\lambda + \nu S -\lambda S) \Rightarrow \nu = \frac{\lambda S |\Delta|}{(\lambda + S)|\Delta| - S}
\end{equation}
So the fixed points (in the phase space $S=1$) are of the form

\begin{equation}
\left\{
\begin{aligned}
& \theta_{\sigma} = \frac{1}{|\Delta|} & ,\mbox{if }\sigma\,\in\,\Delta\\
& \nu_{\sigma} = \frac{\lambda}{|\Delta| \lambda + |\Delta| - 1} & ,\mbox{if }\sigma\,\in\,\Delta\\
& \theta_{\sigma} = \nu_{\sigma} = 0 & ,\mbox{if }\sigma\,\in\,\Omega
\end{aligned}
\right.
\label{eq:fixed-point-voter}
\end{equation}
For the stability analysis, recall that

\begin{equation}
\left\{
\begin{aligned}
& \dot{\theta}_{\sigma} = \theta_{\sigma}\nu - \nu_{\sigma} S \\
& \dot{\nu}_{\sigma} = \lambda \theta_{\sigma} - \nu_{\sigma}(\lambda + S - \theta_{\sigma}) 
\end{aligned}
\right.
\end{equation}
so the Jacobian is given by

\begin{equation}
\left\{
\begin{aligned}
&\frac{\partial \dot{\theta}_{\sigma}}{\partial \theta_{\sigma'}} = \delta_{\sigma, \sigma'}\nu - \nu_{\sigma}\\
&\frac{\partial \dot{\theta}_{\sigma}}{\partial \nu_{\sigma'}} = \theta_{\sigma} - S\delta_{\sigma, \sigma'}\\
&\frac{\partial \dot{\nu}_{\sigma}}{\partial \theta_{\sigma'}} = (\lambda + \nu_{\sigma}) \delta_{\sigma, \sigma'} - \nu_{\sigma}\\
&\frac{\partial \dot{\nu}_{\sigma}}{\partial \nu_{\sigma'}} = -(\lambda + S - \theta_{\sigma})\delta_{\sigma, \sigma'}\\
\end{aligned}\right.
\label{eq:jac}
\end{equation}
It follows that $\mathcal{J}_{\Omega, \Delta} = 0$ and the eigenvalues of $\mathcal{J}$ are obtained joining the ones of $\mathcal{J}_{\Omega, \Omega}$ and $\mathcal{J}_{\Delta, \Delta}$. If $\Omega\neq\varnothing$ then (\ref{eq:fixed-point-voter}) and (\ref{eq:jac}) imply that

\begin{equation}
\mathcal{J}_{\Omega, \Omega} = 
\begin{bmatrix}
\nu \mathds{I} & \,\,- \mathds{I} \\
\lambda \mathds{I} & \,\,-(\lambda + 1) \mathds{I}
\end{bmatrix}
\label{eq:jac-voter-omega}
\end{equation}
so the eigenvalues of $\mathcal{J}_{\Omega, \Omega}$ are the eigenvalues of
\begin{equation}
\begin{bmatrix}
\nu &\,\, - 1 \\
\lambda &\,\, -(\lambda + 1)
\end{bmatrix}
\end{equation}
but this matrix has determinant $-\nu(\lambda + 1) + \lambda$ that equals $\nicefrac{-\nu}{|\Delta|} < 0$ after substituting (\ref{eq:fixed-point-voter}). Since the matrix is real, this means it has two real eigenvalues with different signs. This implies that every fixed point with $\Omega \neq \varnothing$ must be unstable and the only possible stable fixed point is the one where all $M$ opinions are present and have the same proportion. In this case, the Jacobian has the form

\begin{equation}
    \mathcal{J} = \frac{1}{M}
    \begin{bmatrix}
         M\nu \mathds{I} - \nu \mathds{E} &\quad
        -M \mathds{I} + \mathds{E} \\
        (M \lambda + \nu)\mathds{I} - M\nu \mathds{E} & \quad
        -(M \lambda + M - 1)\mathds{I}
    \end{bmatrix}
    \label{eq:jac-voter-coex}
\end{equation}
so the eigenvalues are the ones from the matrices

\begin{equation}
\frac{1}{M}
\begin{bmatrix}
M\nu & -M \\
M \lambda + \nu & -(M \lambda + M - 1)
\end{bmatrix}
\quad\quad\mbox{and}\quad\quad
\frac{1}{M}
\begin{bmatrix}
0 & 0 \\
M \lambda - M^2\nu + \nu & -(M \lambda + M - 1)
\end{bmatrix}
\end{equation}
with multiplicity $M-1$ and 1 respectively. For the first matrix, the trace and determinant are given by

\begin{equation}
    \mathrm{Tr} = \nu - \lambda - \zeta\quad\quad\mbox{and}\quad\quad \mathrm{Det} = -\nu(\lambda + \zeta) + \lambda + \frac{\nu}{M}
\end{equation}
where $\zeta = 1 - \nicefrac{1}{M}$. Substituting (\ref{eq:fixed-point-voter}) this leads to

\begin{equation}
\mathrm{Tr} = \nu - \lambda - \zeta = \frac{\lambda}{\lambda + \zeta} - (\lambda + \zeta) = 
\frac{\lambda(1-2\zeta) - \lambda^2-\zeta^2}{\lambda + \zeta}.
\end{equation}
$\mathrm{Tr}$ is negative for $M>1$, since in this case $\zeta \geq \nicefrac{1}{2} \Rightarrow (1-2\zeta) \leq 0$. For the determinant, note that

\begin{equation}
\frac{\nu}{\lambda} = \frac{1}{\lambda + \zeta} < \frac{1}{\lambda + \zeta - \nicefrac{1}{M}} \Rightarrow \nu\left(\lambda + \zeta - \frac{1} {M}\right) < \lambda \Rightarrow -\nu(\lambda + \zeta) + \lambda + \frac{\nu}{M} > 0 \therefore \mathrm{Det} > 0
\end{equation}
implying that the 2 eigenvalues of this matrix have negative real part. For the second matrix the eigenvalues are trivially 0 (the embedding artifact) and $-(M\lambda + M - 1) < 0$. It follows that the fixed point where all opinions are present in equal proportions is always stable (and is the only stable fixed point).

\section{Observations about the spectra of the matrices found in appendices \ref{ap:sznajd} and \ref{ap:voter}}
\label{ap:extra}

In order to do the linear stability analysis of the Sznajd and voter model, we needed to find the eigenvalues (or at least learn the sign of their real part) of the corresponding Jacobian matrices, namely (\ref{eq:jac-sz-omega}), (\ref{eq:jac-sz-delta}), (\ref{eq:jac-sz-delta-simple}), (\ref{eq:jac-voter-omega}) and (\ref{eq:jac-voter-omega}). Even though these matrices may be arbitrarily large (increasing $M$), the symmetries of the fixed points allow us to find the information we need. All of these matrices can be thought of as special cases of (\ref{eq:jac-sz-delta}), since the spectra of all other matrices follows from setting $a=\ldots=h=0$ in this case. In this extra appendix, we will show that we can get to the conclusion that the spectrum of (\ref{eq:jac-sz-delta}) is obtained from the spectra of (\ref{eq:subjac1}) and (\ref{eq:subjac2}), by making an ansatz about what the eigenvectors are.

We want to find the eigenvalues of the matrix

\begin{equation}
J = 
\begin{bmatrix}
A\mathds{I} + \alpha \mathds{E} & B\mathds{I} + \beta \mathds{E} & a\vec{1} & b\vec{1} \\
C\mathds{I} + \gamma \mathds{E} & D\mathds{I} + \delta \mathds{E} & c\vec{1} & d\vec{1} \\
e\vec{1} & f\vec{1} & x & y \\
g\vec{1} & h\vec{1} & z & w
\end{bmatrix}
\end{equation}
where $\vec{1} = (1,\ldots ,1)\in \mathds{R}^n$, $\mathds{E} = \vec{1}\otimes \vec{1}$ and $\mathds{I}$ is an identity. The first family of eigenvectors is

\begin{equation}
\begin{bmatrix}
r.\vec{t} \,\\
s.\vec{t} \,\\
0 \,\\
0 \,
\end{bmatrix}
\end{equation}
where $\vec{1}.\vec{t} = 0$ (and hence $\mathds{E}.\vec{t} = \vec{0}$). Multiplying $J$ by this we get
\begin{equation}
\begin{bmatrix}
A\mathds{I} + \alpha \mathds{E} & B\mathds{I} + \beta \mathds{E} & a\vec{1} & b\vec{1} \\
C\mathds{I} + \gamma \mathds{E} & D\mathds{I} + \delta \mathds{E} & c\vec{1} & d\vec{1} \\
e\vec{1} & f\vec{1} & x & y \\
g\vec{1} & h\vec{1} & z & w
\end{bmatrix}.
\begin{bmatrix}
r.\vec{t}\, \\
s.\vec{t}\, \\
0\, \\
0\,
\end{bmatrix} = 
\begin{bmatrix}
(Ar+Bs).\vec{t}\, \\
(Cr+Ds).\vec{t}\, \\
0\, \\
0\,
\end{bmatrix}
\end{equation}
so we have and eigenvector with eigenvalue $\mu$ iff

\begin{equation}
\left\{
\begin{array}{l}
Ar+Bs = \mu r \\
Cr+Ds = \mu s
\end{array}
\right.
\end{equation}
that is, iff

\begin{equation}
\begin{bmatrix}
A &B\\
C &D
\end{bmatrix}.
\begin{bmatrix}
r\\
s
\end{bmatrix} = 
\mu \begin{bmatrix}
r\\
s
\end{bmatrix}
\end{equation}
Since the dimension of the space $\left\{\vec{t}\,\,\middle|\,\vec{1}.\vec{t} = 0\right\}$ is $n-1$ it follows that the eigenvalues of 
\begin{equation}
\begin{bmatrix}
A &B\\
C &D
\end{bmatrix}
\end{equation}
show up in the spectrum of $J$ with multiplicity multiplied by $n-1$. The second family of eigenvectors is
\begin{equation}
\begin{bmatrix}
r.\vec{1} \,\\
s.\vec{1} \,\\
t \,\\
u \,
\end{bmatrix}
\end{equation}
Multiplying $J$ by this we get
\begin{equation}
\begin{bmatrix}
A\mathds{I} + \alpha \mathds{E} & B\mathds{I} + \beta \mathds{E} & a\vec{1} & b\vec{1} \\
C\mathds{I} + \gamma \mathds{E} & D\mathds{I} + \delta \mathds{E} & c\vec{1} & d\vec{1} \\
e\vec{1} & f\vec{1} & x & y \\
g\vec{1} & h\vec{1} & z & w
\end{bmatrix}.
\begin{bmatrix}
r.\vec{1} \,\\
s.\vec{1} \,\\
t \,\\
u \,
\end{bmatrix} = 
\begin{bmatrix}
(Ar + \alpha n r + Bs + \beta ns + ant + bnu).\vec{1} \,\\
(Cr + \gamma n r + Ds + \delta ns + cnt + dnu).\vec{1} \,\\
enr + fns + xt + yu \,\\
gnr + hns + zt + wu \,
\end{bmatrix}
\end{equation}
so we have and eigenvector with eigenvalue $\mu$ iff

\begin{equation}
\left\{
\begin{array}{l}
Ar + \alpha n r + Bs + \beta ns + ant + bnu = \mu r \\
Cr + \gamma n r + Ds + \delta ns + cnt + dnu = \mu s \\
enr + fns + xt + yu = \mu t\\
gnr + hns + zt + wu = \mu u
\end{array}
\right.
\end{equation}
that is, iff

\begin{equation}
\begin{bmatrix}
A + \alpha n & B + \beta n & a n & b n \\
C + \gamma n & D + \delta n & c n & d n \\
e & f & x & y \\
g & h & z & w
\end{bmatrix}.
\begin{bmatrix}
r \,\\
s \,\\
t \,\\
u \,
\end{bmatrix} = \mu
\begin{bmatrix}
r \,\\
s \,\\
t \,\\
u \,
\end{bmatrix}
\end{equation}
explaining why the remaining eigenvalues are the ones from the matrix
\begin{equation}
\begin{bmatrix}
A + \alpha n & B + \beta n & a n & b n \\
C + \gamma n & D + \delta n & c n & d n \\
e & f & x & y \\
g & h & z & w
\end{bmatrix}
\end{equation}

Finally, in all cases studied, we end up with one of the eigenvalues equal to 0 (the embedding artifact), which we promptly discard in our analysis. Since this is a fairly niche subject, we offer some explanations here.

In the mean field treatment we did, our variables are the probabilities of finding agents in a given state. This means that the sum of their probabilities
\begin{equation}
    \sum_{\sigma} \theta_{\sigma}
    \label{eq:theta-normalization}
\end{equation}
is a conserved quantity. We could, in principle, use this conserved quantity to eliminate one of the variables, but doing so considerably clouds the symmetries that let us find the fixed points and jacobian spectra. However, the price we must pay to keep the equations in this more symmetric form is that they now describe a dynamics in a space with $D+1$ dimensions, even though the actual phase space we are interested in is a $D$ dimensional embedded submanifold. Because of this, the jacobians will be $D+1\times D+1$ and hence will have an extra eigenvalue that has nothing to do with the dynamics in the manifold (and is hence just an artifact of embedding our phase space in a larger dimensional space). The way to identify this spurious eigenvalue in general is that the corresponding eigenvector is not tangent to the phase space in the fixed point; however, this can be tricky to check. Another way is to write the equations in such a way that the dynamics outside of the phase space move parallel to it. This forces the embedding artifact to be 0, which is usually easier to identify in the final spectrum and also tends to simplify calculations.

In our treatment, this was done by writing the equations in such a way that the quantity in (\ref{eq:theta-normalization}) is a globally conserved quantity. This happens naturally in (\ref{eq:MF-eq-theta}) for the Sznajd model, but is the reason why we don't substitute $S=1$ in (\ref{eq:CM-voter2}) for the voter model. Note that if we did use

\begin{equation}
\dot{\theta}_{\sigma} = \theta_{\sigma}\nu - \nu_{\sigma}
\end{equation}
summing over $\sigma$ this would lead to
\begin{equation}
\dot{S} = \nu (S-1)
\end{equation}
which would lead to an embedding artifact equal to $\nu$. For the coexistence fixed point where $\nu>0$ this wouldn't affect the stability, since it only means that trajectories {\bf outside} of the phase space are being repelled by it. As such, we'd have to identify that this positive eigenvalue is meaningless during the analysis.

\end{document}